\documentclass[]{aa} 
\usepackage{graphicx}
\usepackage{natbib}

\newcommand{\Teff}{\hbox{$T\sb{\rm eff}$}}          
\newcommand{\logg}{\hbox{$\log g$}}
\newcommand{\Msun}{\hbox{M$\sb{\odot}$}}

\newcommand{\Halpha}{\hbox{H$\alpha$}}

\newcommand{\vsini}{\hbox{$v\,\sin\,i$}}

\begin{document}

   \title{Rotation velocities of white dwarfs determined from the Ca\,II~K line
   \thanks{Based in part on data obtained at the Paranal
   Observatory of the European Southern Observatory for programs
   165.H-0588 and 167.D-0407}}

   \author{L. Berger\inst{1}
           \and
           D. Koester\inst{2}
	   \and 
           R. Napiwotzki\inst{3}
	   \and
	   I.~N. Reid  \inst{4}
	   and
	   B. Zuckerman \inst{5}
           }
   \institute{Institut f\"ur Experimentelle und Angewandte Physik,
             University of Kiel, {D-24098 Kiel}, Germany
     \and
   Institut f\"ur Theoretische Physik und Astrophysik, 
             University of Kiel, {D-24098 Kiel}, Germany
     \and
   Centre for Astrophysics Research, STRI,
   University of Hertfordshire, Hatfield AL10 9AB, UK
    \and
Space Telescope Science Institute, 3700 San Martin Drive, Baltimore,
           MD 21218, USA
    \and
Department of Physics \& Astronomy and Center for Astrobiology, 
UCLA, Los Angeles, CA 90095-1562, USA 
}

   \offprints{L. Berger\\ \email{berger@physik.uni-kiel.de}}

   \date{}

\authorrunning{L. Berger et al.}

\titlerunning{Rotation velocities of white dwarfs}

\abstract{We determine projected rotation velocities \vsini\ in DAZ
     white dwarfs, for the first time using the rotational broadening
     of the Ca\,II~K line. The results confirm previous findings that
     white dwarfs are very slow rotators, and set even more stringent
     upper limits of typically less than 10~km/s. The few exceptions
     include 3 stars known or suspected to be variable ZZ Ceti stars,
     where the line broadening is very likely not due to rotation. The
     results demonstrate that the angular momentum of the core cannot
     be preserved completely between main sequence and final stage.
      }

\maketitle

\section{Introduction}
The last years have seen a new generation of stellar evolution models
emerging, which try to incorporate the influence of rotation on
internal structure and evolution \citep[see e.g.][ and later papers of
these groups]{Meynet.Maeder97, Langer.Heger.ea98, Langer98,
Heger.Langer.ea00}. These studies have been motivated by the
realization that stellar surface abundances in many areas of the
Hertzsprung-Russell diagram require mixing processes in addition to
convection. The effects of rotation in massive stars
are described in detail in \cite{Maeder.Meynet00}, which gives also
many relevant references. In the calculation of rotating stellar
models inevitably the question of angular momentum transport arises,
which cannot yet be described self-consistently from first
principles. Ordinary molecular viscosity is much too small, and other
mechanisms have to be invoked, e.g. turbulence created through shear
instability \citep{Spiegel.Zahn70, Zahn74, Zahn75, Zahn92}, generation
of internal gravity waves \citep{Talon.Kumar.ea02, Talon.Charbonnel03,
Talon.Charbonnel04}, or magnetic torques \citep{Spruit98}.

In this situation the observation of white dwarf rotation can provide
some constraints since it tells us the remaining angular momentum of
the typical 0.6~\Msun\ interior core of the progenitor at the end of
stellar evolution. If the total angular momentum of the core is
preserved one would expect equatorial velocities in the range of
15~km/s for solar type slow-rotating progenitors to 2000~km/s for a
fast-rotating A0 star, corresponding to rotation periods from 1~h down
to a few minutes (a number useful to remember is that for a 0.6~\Msun\
white dwarf an equatorial velocity of 15~km/s corresponds to a 1~h
rotation period). 

The only detailed calculation with a sophisticated treatment of
angular momentum transfer, which makes a specific prediction for the
final white dwarf rotational velocity is the work of
\citet{Langer.Heger.ea99}. For the 0.8~\Msun\ white dwarf resulting
from the evolution of a rotating 3~\Msun\ model they expect an
equatorial rotation of 28\,km/s. 

On the other hand -- assuming the core to remain
strongly coupled to the outer expanding envelope in the giant stage --
the rotation of the final white dwarf should be extremely slow, with
periods in the range 30 to 1000 years \citep{Spruit98}. Rotation
periods of hours to days have in a few cases been determined from the
rotational splitting of pulsation frequencies; a list is given in
\cite{Spruit98}, to which could be added a period of 13~hrs for L19-2
and 58~hrs for GD165 \citep{Bradley01}. Periods around one day have
also been measured from variable polarization
\citep{Schmidt.Norsworthy91, Schmidt.Smith95}, but for some of these
magnetic white dwarfs no variation has been observed, implying
rotation periods of at least a century.

The great majority of rotation data for white dwarfs have been
determined spectroscopically, exclusively using the sharp NLTE core of
the \Halpha\ line in DA white dwarfs \citep{Pilachowski.Milkey84,
Pilachowski.Milkey87, Milkey.Pilachowski85, Koester.Herrero88,
Heber.Napiwotzki.ea97, Koester.Dreizler.ea98,
Karl.Napiwotzki.ea05}. The sample now contains about 50 stars with
determined $v \sin i$. A few of these objects show a line broadening
suggesting velocities of a few dozen km/s, but in many cases the
unsatisfactory line fits indicate that the reason is not rotation but
magnetic splitting or unknown, as in the case of the ZZ Ceti stars
\citep{Koester.Dreizler.ea98}. In the majority of objects only upper
limits can be determined which typically range from 10 to 30~km/s and
are thus marginally compatible with the result of
\citet{Langer.Heger.ea99}. While this does not rule out that most
white dwarfs could be extremely slow rotators, \cite{Spruit98}, taking
all available empirical evidence and following his idea of strong
core-envelope coupling felt a need for mechanisms, which would {\em
accelerate} the white dwarfs again to the observed values after the
efficient loss of angular momentum in the giant phase.

There is thus clearly a strong motivation to determine more stringent
upper limits using the spectroscopic method. This has become possible
after the detection of a large number of DAZ objects found in the
studies of \cite{Zuckerman.Reid98, Zuckerman.Koester.ea03} and as a
byproduct of the search for Supernova~Ia progenitors
\citep{Napiwotzki.Christlieb.ea03, Napiwotzki.Christlieb.ea01,
Koester.Rollenhagen.ea05}. The Ca\,II~K line in these hydrogen-rich
white dwarfs is deep and quite narrow and should be useful for
measuring much lower rotation velocities than is possible with
\Halpha. In this paper we report the results of such a study.

\begin{figure}
\includegraphics[width=6.5cm,angle=90]{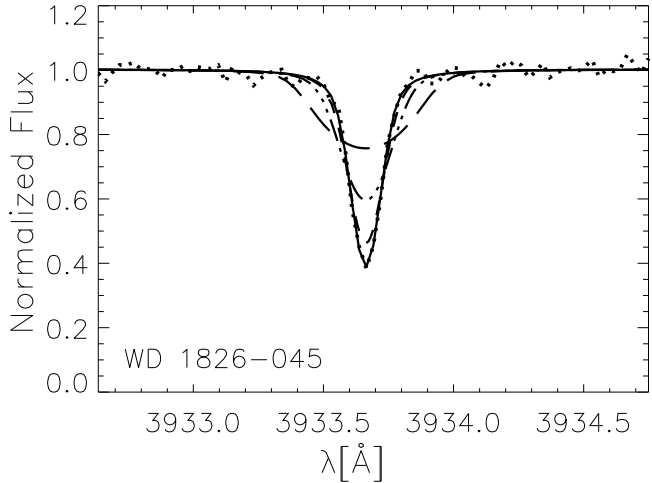}
\caption{Ca\,II~K line profile in WD\,1826-045. Dotted: observed
  profile; continuous: non-rotating model after convolution with
  instrumental profile (falls on top of the observation in the line
  center!); dashed and dot dashed with decreasing line depths:
  rotating models with \vsini\ = 5, 10, 20 km/s
\label{wd1826}}
\end{figure}

\begin{table}[htb!] 
\caption{Atmospheric parameters \Teff\ in K, surface gravity \logg,
  and Ca abundance for the sample. [Ca/H] is the logarithm of the Ca/H
  abundance ratio by numbers. Column Ref gives the source of the
  parameters with Z = \cite{Zuckerman.Koester.ea03}, R =
  \cite{Rollenhagen04} and \cite {Koester.Rollenhagen.ea05}. Columns K
  and S give the number of available spectra from the Keck or the SPY
  samples. {\em Please note that the Ca abundance for HS\,0047+1903 in 
\cite {Koester.Rollenhagen.ea05} was based on an erroneous equivalent
  width. The correct abundance is given in Table~\ref{tabres}}
\label{para}}
\centering\tt
\begin{tabular}{lrrrrrr} 
\hline
Object & $T_{eff}$ & $\log\,g$ & [Ca/H]& Ref & K & S \\
\hline
WD\,0032-175   &  9235 & 8.0 & -10.2 & Z & 1 & \\
HS\,0047+1903  & 16600 & 7.8 &       & R &   & 1 \\  
HE\,0106-3253  & 15700 & 8.0 &  -6.4 & R &   & 2 \\
WD\,0208+296   &  7201 & 7.9 &  -8.8 & Z & 2 & \\
WD\,0235+064   & 11420 & 7.9 &  -9.0 & Z & 1 & \\
WD\,0243-026   &  6798 & 8.2 &  -9.9 & Z & 1 & 2\\
WD\,0245+541   &  5190 & 8.2 & -12.7 & Z & 1 &\\
HS\,0307+0746  & 10200 & 8.1 &  -7.6 & R &   & 1 \\
WD\,0408-041   & 14400 & 7.8 &  -7.1 & R &   & 2\\
WD\,0543+579   &  8142 & 8.0 & -10.3 & Z & 3 & \\
WD\,0846+346   &  7373 & 8.0 &  -9.4 & Z & 2 & \\
WD\,1015+161   & 19300 & 7.9 &  -6.3 & R &   & 2 \\
WD\,1102-183   &  8026 & 8.0 & -10.4 & Z & 1 & \\
WD\,1116+026   & 12200 & 7.9 &  -7.3 & R &   & 2 \\
WD\,1124-293   &  9680 & 8.0 &  -8.5 & Z & 2 & 2 \\
WD\,1150-153   & 12800 & 7.8 &  -6.7 & R &   & 2 \\
WD\,1202-232   &  8619 & 8.0 &  -9.8 & Z & 2 & 2 \\
WD\,1204-136   & 11468 & 8.0 &  -7.7 & Z & 1 & 2 \\
WD\,1208+576   &  5830 & 7.9 & -11.0 & Z & 1 & \\
HE\,1225+0038  &  9400 & 8.1 &  -9.7 & R &   & 3 \\
WD\,1257+278   &  8481 & 7.9 &  -8.1 & Z & 2 & \\
HE\,1315-1105  &  9400 & 8.4 &  -9.2 & R &   & 2 \\
WD\,1337+705   & 20435 & 7.9 &  -6.7 & Z & 3 & \\
WD\,1344+106   &  6945 & 8.0 & -11.1 & Z & 2 & \\
WD\,1407+425   &  9856 & 8.0 &  -9.9 & Z & 1 & \\
WD\,1455+298   &  7366 & 7.6 &  -9.3 & Z & 2 & \\
WD\,1457-086   & 20400 & 8.0 &  -6.3 & R &   & 2 \\
WD\,1614+160   & 17400 & 7.8 &  -7.2 & R &   & 2 \\ 
WD\,1633+433   &  6569 & 8.1 &  -8.6 & Z & 2 & \\
WD\,1821-131   &  7029 & 8.4 & -10.7 & Z & 1 & \\
WD\,1826-045   &  9480 & 7.9 &  -8.8 & Z & 1 & 1 \\
WD\,1858+393   &  9470 & 8.0 &  -7.8 & Z & 2 & \\
WD\,2115-560   &  9700 & 8.1 &  -7.6 & R &   & 1 \\
HS\,2132+0941  & 13200 & 7.7 &  -7.7 & R &   & 2 \\
WD\,2149+021   & 17300 & 7.9 &  -7.6 & R &   & 2 \\
HE\,2221-1630  & 10100 & 8.2 &  -7.6 & R &   & 2 \\
HS\,2229+2335  & 18600 & 7.9 &  -6.3 & R &   & 2 \\
WD\,2326+049   & 11600 & 8.1 &  -6.9 & Z & 5 & 2 \\
\hline
\end{tabular}
\end{table}

\section{Observations} \label{observations}
The sample consisted of spectra for 38 single cool DA white dwarfs
coming from two different sources. The first is the SPY survey (ESO
Supernova Ia Progenitor Survey) for double-degenerates.  In the search
for variable radial velocities a large number of high-resolution
spectra are obtained with the UVES spectrograph at the ESO VLT Paranal
Observatory, which are useful also for many other
projects. \cite{Rollenhagen04} and \cite{Koester.Rollenhagen.ea05} used this
sample to identify hydrogen-rich white dwarfs with Ca\,II lines and
those objects form the first part of our present sample (excluding
those where the lines are interstellar). Details of the observations
and the reduction procedures are described in
\cite{Napiwotzki.Christlieb.ea03} and \cite{Koester.Napiwotzki.ea01};
we therefore repeat here only the facts relevant for this project. The
observations used a wide slit of 2\farcs 1, because the SPY was
conducted as a filler project for bad weather conditions. This is
important since the nominal resolution with this slit is 18500
corresponding to 0.21~\AA\ at the Ca\,II~K line, however, the actual
resolution can be higher if the seeing disk is smaller than the
slit. The actual seeing during the SPY observations typically varied
between 1\farcs0 to 1\farcs5.

The second source of sample objects are Keck observations of DAs in a
search for DAZs with metals. The observations are described in
\cite{Zuckerman.Reid98} and \cite{Zuckerman.Koester.ea03}. Here the
nominal resolution with a slit width of 1\farcs 15 is 34000,
corresponding to 0.12~\AA\ at Ca\,II~K, but also in this case the actual
resolution might be slightly higher at times of excellent seeing.

For most objects more than one spectrum is available, for some even
spectra from both surveys.  Atmospheric parameters \Teff, \logg, and
Ca abundances for the sample are given in
\cite{Koester.Rollenhagen.ea05} and \cite{Zuckerman.Koester.ea03}
(which also show some representative spectra) and are collected in
Table~\ref{para}.  For our analysis the spectra were shifted slightly
to bring the K line in agreement with the laboratory wavelength and
then normalized to a continuum of 1.0 with a straight line fitted to
the continuum approximately 5~\AA\ on both sides of the line center.

\section{Determination of rotation velocities}
Rotation velocities were determined using theoretical model
atmospheres with spectral lines broadened by a rotational broadening
function \citep[see e.g.][]{Unsold68}. Our model grid is calculated
assuming LTE (local thermodynamic equilibrium) with methods and input
physics described in \cite{Finley.Koester.ea97},
\cite{Koester.Wolff00}, and \cite{Homeier.Koester.ea98}. The Ca\,II
lines have the advantage compared to the Balmer lines -- the only
lines used in the past for spectroscopic determinations in white
dwarfs -- that the central Doppler core from thermal broadening is
much narrower because of the higher atomic weight.

On the other hand, the Balmer lines originate from the dominant
element in DAs and therefore only two parameters, \Teff\ and \logg,
are needed to describe the fitting model. In the case of the K line an
additional parameter is the Ca abundance.  The results given in
\cite{Zuckerman.Koester.ea03} and \cite{Koester.Rollenhagen.ea05} have
been determined using interpolations in pre-calculated tables of
equivalent widths. In the calculations of the theoretical equivalent
widths a fixed interval was assumed, and no attempt was made to match
this exactly in the observed spectra; moreover, the spacing of the
grid of abundances was rather coarse. It is therefore not surprising
that calculated spectra for the ``exact'' atmospheric parameters using
these Ca abundances did not always match the spectra perfectly. We
have then changed the Ca abundance (typically less than a factor of
two) until the EW of the observation agreed exactly with that of the
theoretical model. This is obviously a prerequisite for the
determination of rotational broadening of the profile.

The standard rotational broadening profile uses two free parameters,
the limb darkening coefficient $\beta$ and the projected equatorial
rotation velocity \vsini. Here $i$ is the inclination of the rotation
axis against the line-of-sight, ranging from 0 to 90 degrees (note
that only this combination can be determined with the spectroscopic
method). Our theoretical models show that $\beta$ varies from
$\approx 0.2$ at the center of the K line to $\approx 1.2$ in the far
wing. Since the parameter has almost no influence on the result we
have continued to use $\beta = 0.15$ which is used by all authors for
\Halpha. The best fit value for \vsini\ and the errors were then
determined with a $\chi^2$ fitting routine very similar to that used
in \cite{Koester.Dreizler.ea98}.  Figure~\ref{wd1826} shows the
observed line in WD\,1826-045 as a typical example together with
theoretical models with different rotational velocities applied. The
figure demonstrates that rotational velocities of a few km/s can
easily be measured from the Ca line.

\begin{table}
\caption{Rotational velocities. When several spectra of the same
  object are of comparable quality, the results do not differ
  significantly and we therefore give only the result from the best
  spectrum. The only exception is WD$2326+049$, a known ZZ Ceti
  variable. [Ca/H] is the Ca
  abundance as determined from fitting the line profile, which is
  sometimes slightly different from that in the original source (see
  Table~\ref{para}; the corrected EW for HS\,0047+1903 used for the
  abundance in this table is 77 m\AA). $v_2 = v\,\sin\,i$ is the rotation
  velocity determined using either the nominal resolution of the
  spectrographs (34000 for Keck, 18500 for SPY), or a slightly higher
  resolution if the line depth cannot be reproduced at nominal
  resolution. In the latter case the values for the resolution are
  given in column Res. Errors are $1\,\sigma$
  errors. $v_1 = (v\,\sin\,i)_{max}$ is an upper limit using the optimum
  resolution as explained in the text
\label{tabres}}
\centering \tt
\begin{tabular}{lrrrr}
\hline
Object &  [Ca/H] & $v_1$ & $v_2$ & Res \\
\hline
WD\,0032-175 & -10.5 & 3.8 & 0.0 $\pm$ 2.8 & 46300 \\
HS\,0047+1903& -6.1  & 11.6& 9.9 $\pm$ 6.1 &       \\
HE\,0106-3253& -5.8 & 6.1 & 0.0 $\pm$ 4.9 & 28100 \\
WD\,0235+064 & -9.2 & 5.1 & 0.0 $\pm$ 4.0 & 39300 \\
WD\,0243-026 & -9.8 & 4.7 & 2.5 $\pm$ 4.2 & 32800 \\
WD\,0245+541 & -11.7 & 0.0 & 0.0 $\pm$ 6.6 & \\
HS\,0307+0746& -7.1 & 9.1 & 5.4 $\pm$ 5.5 & 24600 \\
WD\,0408-041 & -6.6 & 7.1 & 0.0 $\pm$ 5.1 & 24600 \\
WD\,0543+579 & -10.5 & 5.0 & 3.4 $\pm$ 1.7 & 46300 \\
WD\,1015+161 & -5.9 & 10.1 & 2.4 $\pm$ 4.7 & 20700 \\
WD\,1102-183 & -10.3 & 8.1 & 5.9 $\pm$ 1.9 & \\
WD\,1116+026 & -6.5  & 11.4& 3.6 $\pm$ 3.8 & \\
WD\,1124-293 & -8.2 & 5.8 & 0.0 $\pm$ 2.2 & \\
WD\,1150-153 & -6.0 & 14.3 & 9.5 $\pm$ 5.4 & \\
WD\,1202-232 & -9.8 & 5.8 & 0.0 $\pm$ 0.9 & \\
WD\,1204-136 & -7.2 & 11.1 & 9.7 $\pm$ 0.8 & \\
WD\,1208+576 &-10.8 &  8.8 & 6.5 $\pm$ 1.9 & \\
HE\,1225+0038& -9.7 & 9.4 & 0.0 $\pm$ 7.0 & 21300\\
HE\,1315-1105& -9.4 & 6.7 & 1.5 $\pm$ 4.8 & 27100 \\
WD\,1337+705 & -6.7 & 4.4 & 0.0 $\pm$ 1.2 & 41400 \\
WD\,1344+106 & -11.3 & 7.2 & 4.0 $\pm$ 2.5 &  \\
WD\,1407+425 & -9.8 & 8.9 & 6.7 $\pm$ 1.5 & \\
WD\,1457-086 & -6.1 & 9.6 & 0.0 $\pm$ 8.5 & 21900 \\
WD\,1614+160 & -7.2 & 2.2 & 0.4 $\pm$ 4.9 & 35750 \\
WD\,1821-131 & -10.9 & 18.0 & 15.8 $\pm$ 4.6 & \\
WD\,1826-045 & -8.6 & 3.1 & 0.0 $\pm$ 1.3 & 49200 \\
HS\,2132+0941& -7.1 & 8.5 & 0.0 $\pm$ 8.3 & 21260 \\
WD\,2149+021 & -7.7 & 10.7& 0.0 $\pm$ 10.3& 20700 \\
HE\,2221-1630& -7.2 & 4.2 & 2.1 $\pm$ 4.0 & 34200 \\
HS\,2229+2335& -5.9 & 6.5 & 0.0 $\pm$ 5.2 & 27100 \\
WD\,2326+049 & -6.4 & 20.5 & 19.8 $\pm$ 0.5 & \\
             & -6.3 & 15.7 & 14.6 $\pm$ 0.5 & \\
             & -6.3 & 28.2 & 27.6 $\pm$ 0.4 & \\
             & -6.3 & 26.9 & 26.2 $\pm$ 0.3 & \\
             & -6.3 & 28.4 & 28.0 $\pm$ 0.4 & \\
             & -6.4 & 15.3 & 10.7 $\pm$ 2.1 & \\
             & -6.4 & 22.5 & 20.0 $\pm$ 0.9 & \\
\hline
\end{tabular}
\end{table}

\begin{figure}
\includegraphics[width=6.0cm,angle=90]{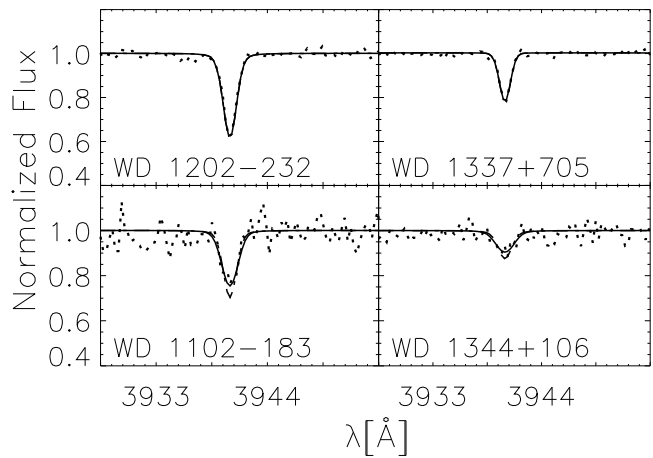}
\caption{Typical examples for the determination of rotational
  velocities for the Ca\,II~K line, ranging from some of the best
  spectra to marginal. Dotted: observed
  profile; dashed: non-rotating model after convolution with
  instrumental profile; continuous: best fit with rotation.
  In the two top figures the best fit is zero rotation and the models
  fall on top of each other.
\label{goodrot}}
\end{figure}

\subsection{A remark on the standard spectroscopic method}
Assuming that the intensity of the radiation field at the surface of a
star depends only on wavelength $\lambda$ and the angle $\theta$
against the normal, the observational quantity for a non-rotating star
is the disk-averaged intensity
\[ \bar{I} = \frac{1}{\pi}\,\int \int I(\mu,\lambda)\,dx dy
\]
where $\mu = \cos \theta$, and $x,y$ are Cartesian coordinates on the
visible surface of the star in units of the stellar radius $R$.

In the case of a rotating star we have in addition a Doppler shift,
which depends on the position on the disk
\[ \bar{I} = \frac{1}{\pi}\,\int\limits_{-1}^{+1} 
      \int\limits_{-\sqrt{1-x^2}}^{+\sqrt{1-x^2}}
      I(\mu,\lambda-\Delta\lambda(x,y))\,dx dy 
\]
If the $y$-axis is chosen parallel to the rotation axis,
$\Delta\lambda$ will depend on $x$ only.  This integral can easily be
evaluated numerically if theoretical models are available for the
intensity as a function of $\mu$. It is common practice, however,
to proceed with an analytic evaluation by assuming a limb-darkening
law
\[  I(\mu) = I_c ((1-\epsilon) + \epsilon \mu)
\]
with central intensity $I_c$. This law can also be written relative to
the intensity at the limb $I_l$
\[ I(\mu) = I_l (1 + \beta \mu)
\]
with the limb-darkening coefficient $\beta = \epsilon/(1-\epsilon)$
used above.
Introducing this into the integration over the disk, with $\mu$ a
function of position $x, y$, the integration can be transformed into
a convolution of the disk-averaged intensity for a non-rotating star
\[ \bar{I}_{rot} = 
       \int\limits_{-1}^{+1} G(x) \bar{I}(\lambda - x \Delta\lambda_l)
       \,dx
\]
with the maximum shift at the limb $\Delta\lambda_l$, and
\[ G(x) = \frac{4 (1-\epsilon) + \epsilon \sqrt{1-x^2} + \pi \epsilon
  (1-x^2)}{2 \pi (1 - \epsilon/3)}
\]
\cite[see e.g.][]{Unsold68}. This is a large simplification, since
only the theoretical disk-averaged intensity (also called the
astrophysical flux) needs to be calculated, which is always needed for
the analysis of spectra, rotating or not. However, the derivation
makes the assumption that the limb darkening law is independent of
wavelength, in particular that the coefficients $\epsilon$ or $\beta$
are constant over the line profile. This is clearly not the case for
the Ca\,II~K line as mentioned above and it is also not true for
\Halpha. We have therefore studied the effect of this assumption by
comparing the exact direct integration with angle-dependent intensities with
the  simplified method using the convolution technique. While this has
been studied before in the literature for other types of stars it has
to our knowledge never been tested for white dwarf spectra.

The result of this test is \citep{Pruin05} that small differences
between the two methods exist in the line cores, but that the error
made when using the simple convolution method is much smaller than
1~km/s, and thus negligible in our study.

\begin{figure}[ht]
\includegraphics[width=21.3cm,angle=90]{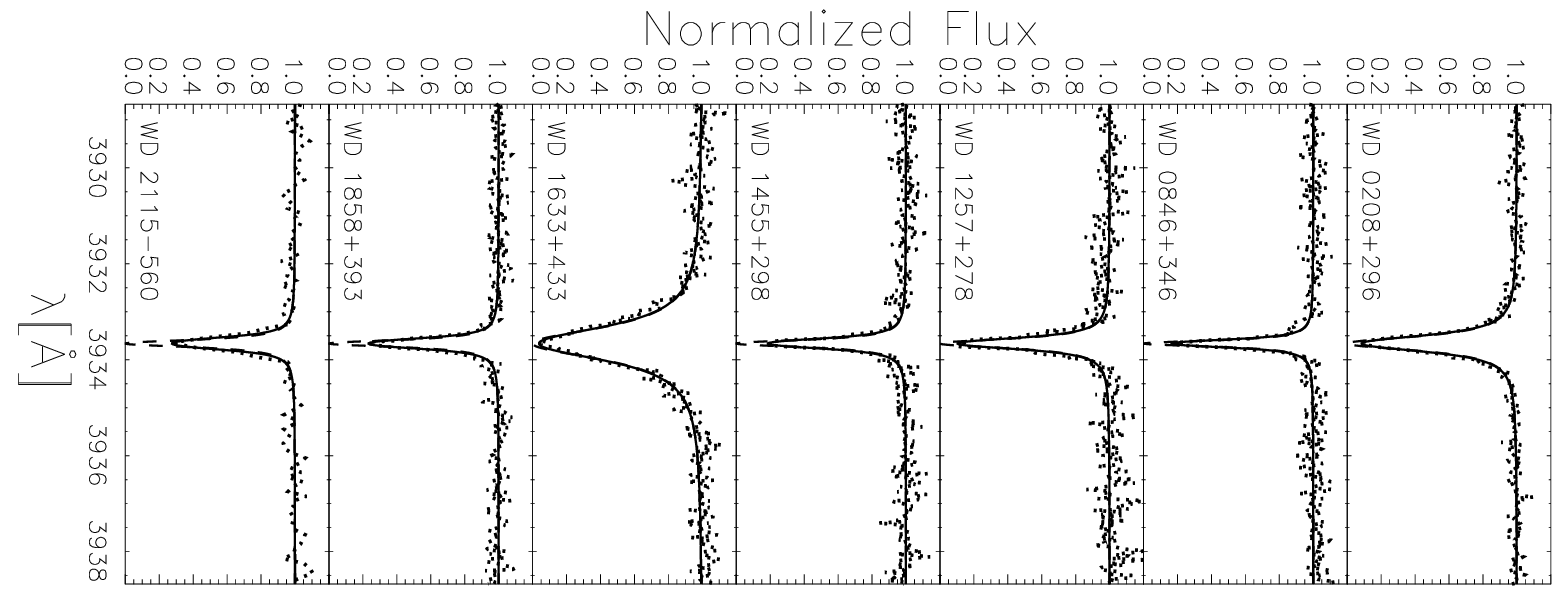}
\caption{Seven objects fitted with Lorentz profiles. See text
\label{deep}}
\end{figure}

\begin{table}
\caption{Upper limit to the rotational velocities for the objects with
  deeper observed line profiles than is possible with a non-rotating
  model.  The limits were determined using a Lorentz line profile with
  central intensity zero and equivalent width (EW, in m\AA) equal to
  the observed spectrum
\label{ressix}}
\centering\tt
\begin{tabular}{lcrr} 
\hline
Object & $S/N$ & EW & $(v\,\sin\,i)_{max}$ \\
\hline
WD\,0208+296 & 37 & 534  & 8\\
WD\,0846+346 & 24 & 245  & 6\\
WD\,1257+278 & 15 & 375  & 7\\
WD\,1455+298 & 21 & 282  & 8\\
WD\,1633+433 & 19 & 1200 & 13\\
WD\,1858+393 & 27 & 285  & 9\\
WD\,2115-560 & 24 & 292  & 11 \\
\hline
\end{tabular}
\end{table}

\subsection{Observed and theoretical line profiles}
When comparing the observed and theoretical profiles after taking into
account instrumental broadening with the nominal resolution in the
latter, we obtained nearly perfect fits for about half of the
objects {\em without rotational broadening}, implying zero or at least
extremely low rotation. In the other half of objects the observed line
profile had a deeper core than our ``unrotated'' models. Since
rotational broadening can only decrease the central depths, we could
thus not obtain a fit. 

What could be the reason for this discrepancy? The line profiles are
of course influenced by the reduction process, e.g. the subtraction of
sky background, to mention just one step which could affect especially
the line cores. Indeed, when we have several spectra of the same
object, there are sometimes differences in equivalent width and also
line depth. Another possibility is on the modeling side: the largest
discrepancies (in seven objects) occur for very deep lines, where the
core is saturated. In these cases the line core originates in the
outermost layers of the atmosphere and it is quite possible that NLTE
effects (not included in our models) will make the model lines deeper,
as is well known for \Halpha.

We believe, however, that in most objects the reason is that the
actual resolution of the observation was better than nominal. This is
supported by the finding that in several spectra of the same stars the
equivalent width is constant but the line profile changes, with the
line being deeper and narrower in some of the spectra. This strongly
indicates a changing actual resolution for periods with better
seeing. 

Unfortunately we do not know the real resolution for the
observation. In order to be able to nevertheless extract some
information on the rotation velocities we have therefore increased the
resolution (narrowed the instrumental profile) until our models
achieved a good fit without rotation. We have then used our $\chi^2$
routine to determine \vsini. Obviously the result must be compatible
with zero, but we nevertheless obtain a reasonable upper limit from
the errors provided by the fit. The results obtained for 31 objects
for which this procedure led to a good fit with the nominal or within
a plausible range of resolutions (see below) are shown in
Table~\ref{tabres}. A number in the resolution column indicates the
higher than nominal resolution used in the fit.

There is of course some arbitrariness involved with this method, since
the resolution could actually be even higher, and could also have been
higher in the cases where our zero-rotation models had lines deep
enough. In order to analyze the uncertainty we have assumed a maximum
plausible resolution of 39500 and 65500, equivalent to seeing of
$1\farcs 0$ and $0\farcs 6$, respectively, for SPY and Keck
spectra. The column \vsini (max) gives the results of the fitting
procedure, which should be considered as very firm upper limits.

Fig.\ref{goodrot} presents four typical examples from spectra with
high to rather low signal-to-noise ratios.

\subsection{Error estimates}
We estimated the errors caused by the uncertainties of $T_{eff}$,
$log\,g$, [Ca/H], $\beta$, and the resolution. The summed
errors from the first four quantities were of the same order of
magnitude as the formal $1\sigma$-error of the $\chi^{2}$-fitting
given in Table~\ref{tabres}. The greatest error was caused by
the uncertainty of the resolution and can be estimated by comparing
the values in the column \vsini\ and $(\vsini)_{max}$.

\subsection{Objects with very deep line cores}
Even at the assumed maximum increase of the nominal resolution there
remain seven objects which cannot be fitted by unrotated profiles. All
of these objects have strong Ca lines, which are saturated in the
core. In these cases we have used a Lorentz profile with zero central
intensity and a width calculated to reproduce the observed EW. These
profiles were then used to fit the observed profiles, resulting in
firm upper limits to the rotational velocities. These results are
given in Table~\ref{ressix}. The spectra and fits are shown in
Fig.~\ref{deep}, where the dotted lines are the observations, the
dashed lines Lorentz profiles with zero central intensity as an
extreme simulation of the line profile, and the continuous lines are
the ``rotated'' profiles.

\section{Discussion}
21 objects in Table~\ref{tabres} and all 7 in Table~\ref{ressix} are
compatible with zero rotation velocities (within $1\sigma$
errors). Even allowing for a higher than nominal resolution, within
the plausible range, gives upper limits typically smaller than
10~km/s. This is a clear improvement compared to the spectroscopic
determination from the core of \Halpha, with many upper limits in the
range 10 - 30~km/s \citep{Koester.Dreizler.ea98,
Karl.Napiwotzki.ea05}. Unfortunately we are aware of only two \Halpha\
determinations for objects of our sample, both in
\cite{Koester.Dreizler.ea98}. For WD\,2115$-$560 the upper limit from
\Halpha\ is 35~km/s, while we obtain 11~km/s from the Ca line. 
The other common object is the ZZ Ceti star WD\,2326$+$049
(G29-38). From \Halpha\ a velocity of $45\pm 5$ km/s is obtained, but
the authors note that the fit is not good and the broadening very
likely not due to rotation. From our Ca determination we find a range
of 14-28~km/s from different spectra. 

G29-38 has been very extensively studied as a pulsator, both with
photometry and spectroscopy. \cite{van-Kerkwijk.Clemens.ea00} detected
periodic Doppler shifts in the line profiles with amplitudes up to 5
km/s, which they tentatively associated with the horizontal motions of
g-mode oscillations. \cite{Thompson.Clemens.ea03} recently found
Doppler shifts in G29-38 corresponding to radial velocity changes up
to 16.5~km/s. By combining a large number of spectra with close to
zero Doppler shift and comparing with an average spectrum over several
periods they found no significant difference in the line
profiles. They concluded that pulsation cannot be the reason for the
flat-bottom line profiles. This conclusion may be somewhat premature,
however, since the study only demonstrates that even the profile in
one particular pulsation phase does not fit a rotationally broadened
model spectrum. It is plausible that even with an average shift of
zero velocity in that phase different parts of the stellar surface may
show positive as well as negative velocity along the line-of-sight,
leading to zero shift but a broadening of the profile. This can only
be decided with numerical simulations of the pulsating surface
including all shifts and realistic center-to-limb variations.  In any
case the difference between the \Halpha\ and Ca results and the
significant variation of the Ca ``rotation velocities'' confirm
convincingly that rotation is not the origin of the broadening.

WD\,2326$+$049 is one of the 10 objects for which we determine a
non-zero rotation velocity for the best fitting model. Two others are
WD\,1150-153 and WD\,1204-136, which according to their atmospheric
parameters fall within or near the ZZ Ceti instability strip.
WD\,1204-136 was found to be non-variable (upper limit 0.3\%) by
\citet{Bergeron.Fontaine.ea04}, while WD\,1150-153 to our knowledge
has not been observed for variability. The latter is on our candidate
list for upcoming time-resolved photometry, but in any case it seems
very likely that the same mechanism -- as yet unidentified -- is
responsible for the line shapes as in the known ZZ Ceti stars
\citep{Koester.Dreizler.ea98}. Four other white dwarfs with non-zero
rotation (WD\,0543$+$579, WD\,1407$+$425, WD\,1344$+$106,
WD\,1821$-$131) are cool and the parameters are based on photometry
only. In the first two cases the parallax is not known and thus the
surface gravity cannot be determined and was fixed at 8.0
\citep{Zuckerman.Koester.ea03}.  However, even in these cases the
rotation velocities are definitely very small, $< 18$ km/s at most.

The results of this paper confirm the findings of all previous
studies, that white dwarfs are slow rotators, and set even more
stringent limits for those objects with Ca lines. Typical rotation
velocities are below 10~km/s and may even be close to zero velocity
for many objects. While this could be just marginally consistent with
conservative evolution for solar-type slow-rotating progenitors, the
result would definitely exclude conservation of angular momentum
for more massive progenitors. How many of such white dwarfs could we
expect in our sample?

Assuming for simplicity a unique initial-final mass relation between
remnant and progenitor, we find from \cite{Weidemann00} that a
0.6~\Msun\ white dwarf originates from a 2~\Msun\ main sequence
progenitor. 0.6~\Msun\ (at \Teff\ = 10000~K) correspond to \logg\ =
8.0. Taking the numbers for the atmospheric parameters in
Table~\ref{para} at face value, we have 27 objects with a \logg\
determination (versus {\em assuming} 8.00), of which 11 have \logg\
$\ge 8.0$ and should have originated from fast-rotating main sequence
stars more massive than 2.0~\Msun. While the exact numbers may be
uncertain, there is no doubt that these results mean that angular
momentum in the core cannot be preserved. These results also clearly rule out
the prediction by \cite{Langer.Heger.ea99}, which were based on
stellar model calculations without magnetic fields.

While the original study of \cite{Spruit98} predicted very efficient
angular momentum transport through the magnetic field coupling of core
and outer envelope, the new quantitative formulation of magnetic field
generation and magnetic field induced angular momentum transport
developed in \citet{Spruit02} leads to a more limited angular momentum
transport efficiency. First calculations for massive stars
\citep{Heger.Woosley.ea05} including the \citet{Spruit02} magnetic
field physics show these stars evolving far from rigid
rotation. Similar results are expected for low mass stars (Langer
2005, priv. comm.), which would likely bring the predictions
for the white dwarf rotation velocities in agreement with the new
observations.

Our method could in principle be improved further,
if that is necessary to distinguish between theoretical assumptions.  
The main reason for the current uncertainty is the unknown 
effective resolution due to rather large slits used in the
spectroscopy. With a narrower slit and well controlled resolution an
accuracy of 1-2~km/s could be reached, but the observations would need
a significant amount of observing time on a very large telescope.

\end{document}